\documentclass[prl,aps,twocolumn,superscriptaddress,floatfix]{revtex4-1}

\usepackage{amssymb}
\usepackage{epsfig}
\usepackage{amsmath}
\usepackage{subfigure}
\usepackage{graphicx}
\usepackage{textcomp}
\usepackage{url}
\usepackage{float}
\usepackage{color}
\usepackage{cases}
\usepackage[normalem]{ulem}

\usepackage[titletoc,title]{appendix}

\usepackage[colorlinks=true, urlcolor=blue, anchorcolor=blue, citecolor=blue,filecolor=blue,linkcolor=blue,menucolor=blue
]{hyperref}

\usepackage{color}

\def\Xint#1{\mathchoice
   {\XXint\displaystyle\textstyle{#1}}%
   {\XXint\textstyle\scriptstyle{#1}}%
   {\XXint\scriptstyle\scriptscriptstyle{#1}}%
   {\XXint\scriptscriptstyle\scriptscriptstyle{#1}}%
   \!\int}
\def\XXint#1#2#3{{\setbox0=\hbox{$#1{#2#3}{\int}$}
     \vcenter{\hbox{$#2#3$}}\kern-.5\wd0}}

\def\dashint{\Xint-}

\newcommand{\nn}{\nonumber}
\newcommand{\be}{\begin{equation}}
\newcommand{\ee}{\end{equation}}
\newcommand{\bea}{\begin{eqnarray}}
\newcommand{\eea}{\end{eqnarray}}

\usepackage{fixmath}
\newcommand{\vect}[1]{\mathbold {#1}} 

\begin{document}

\title{Inverse Scattering Method Solves the Problem of Full Statistics of Nonstationary Heat Transfer in the Kipnis-Marchioro-Presutti Model}
\author{Eldad Bettelheim}
\email{eldad.bettelheim@mail.huji.ac.il}
\affiliation{Racah Institute of Physics, Hebrew University of
Jerusalem, Jerusalem 91904, Israel}
\author{Naftali R. Smith}
\email{naftalismith@gmail.com}
\affiliation{Laboratoire de Physique de l'\'{E}cole Normale Sup\'erieure, CNRS, ENS \& Universit\'e PSL, Sorbonne Universit\'e, Universit\'e de Paris, 75005 Paris, France} 
\affiliation{Department of Solar Energy and Environmental Physics, Blaustein Institutes for Desert Research, Ben-Gurion University of the Negev, Sede Boqer Campus, 8499000, Israel}
\author{Baruch Meerson}
\email{meerson@mail.huji.ac.il}
\affiliation{Racah Institute of Physics, Hebrew University of
Jerusalem, Jerusalem 91904, Israel}

\begin{abstract}
We determine the full statistics of nonstationary heat transfer in the Kipnis-Marchioro-Presutti lattice gas model at long times by uncovering and exploiting complete  integrability of the underlying equations of the macroscopic fluctuation theory. These equations are closely related to the  derivative nonlinear Schr\"{o}dinger  equation (DNLS), and we solve them by the Zakharov-Shabat inverse scattering method (ISM) adapted by
Kaup and Newell (1978) for the DNLS. 
We obtain explicit results for the exact large deviation function of the transferred heat for an initially localized heat pulse, where we uncover a nontrivial symmetry relation.

\end{abstract}
\maketitle

\textit{Introduction.} -- Full statistics of currents of matter or energy in macroscopic systems
away from thermodynamic equilibrium is a fundamental quantity that has attracted much attention from statistical physicists in the past two decades.  Major progress has
been achieved  in determining this quantity for nonequilibrium steady states in simple models of interacting particles \cite{Derrida2007, BlytheEvans, AppertRolland,Lecomte}.
Nonstationary fluctuations of current, however, proved to be much harder for analysis \cite{DG2009a,DG2009b,KrMe,MS2013,MS2014,VMS2014,ZarfatyM}.

A convenient and widely used family of models for studying the full statistics of currents is stochastic lattice gases \cite{Spohn,Liggett,KL,Krapivskybook}. One important example is the  Kipnis-Marchioro-Presutti (KMP) model of heat transfer. The KMP model involves
immobile particles occupying a whole lattice and carrying continuous amounts of energy. At each random move the total
energy of a randomly chosen pair of
nearest neighbors is randomly redistributed among them according to uniform distribution.
The KMP model originally
attracted much interest as the first model for which Fourier's law
of heat diffusion at a coarse-grained level was proven rigorously \cite{KMP}. By
now it has become a paradigmatic model of nonequilibrium fluctuations of transport \cite{Bertini2005,BGL,BodineauDerrida,DG2009b,Lecomte,Tailleur,HurtadoGarrido,KrMe,
Pradosetal,MS2013,Peletier,ZarfatyM,Spielberg,Guttierez,Frassek,Benichou}.

\begin{figure}[ht]
\includegraphics[width=0.30\textwidth,clip=]{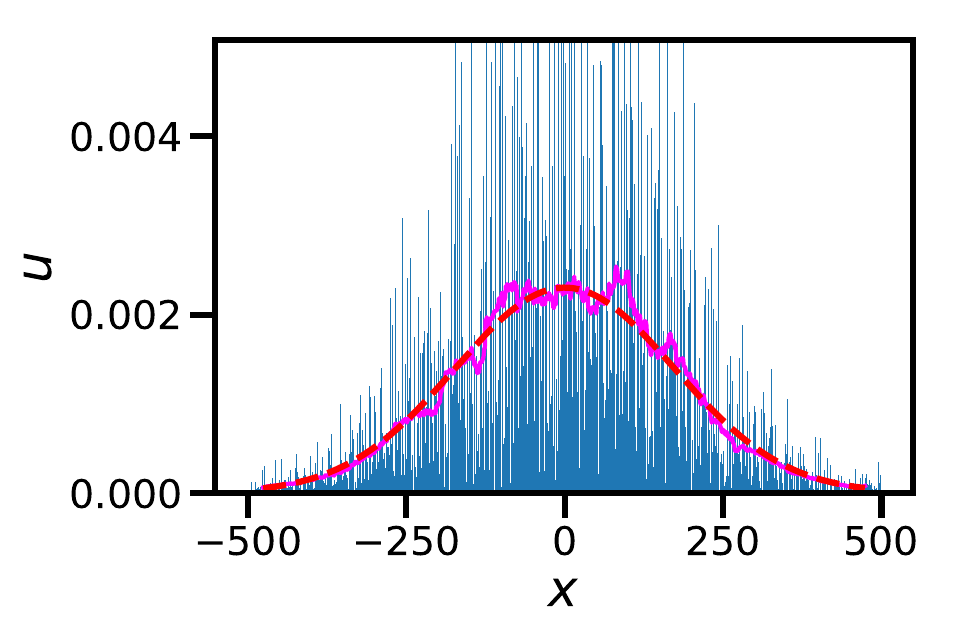}
\caption{Monte-Carlo simulation of the KMP model with $W=1$. Plotted is the simulated temperature profile $u$ as a function of $x$ at time $t=1.5 \times {10}^4$ (bars), its spatial average over each 50 consecutive lattice sites (solid line) and the theoretical Gaussian profile~(\ref{meanfield}) (dashed line).}
\label{fig:sim}
\end{figure}

Here we study a full non-stationary heat-transfer statistics in the KMP model on an infinite one-dimensional lattice. Suppose that only one particle has a nonzero energy 
at $t=0$. Due to the energy exchange with the neighbors, the energy will start spreading throughout the system. At times much longer than the inverse rate of the energy exchange between the two neighbors (equal to $1/2$),  and at distances much larger than the lattice constant (equal to $1$), the mean coarse-grained temperature $\bar{u}(x,t)$ in the KMP model
is governed by the heat diffusion equation \cite{KMP,Spohn,KL} $\partial_t \bar{u} (x,t) = \partial_x^2 \bar{u}(x,t)$.
The initial temperature is a delta-function, $\bar{u} (x,t=0) = W \delta(x)$, and so the solution is
\begin{equation}\label{meanfield}
\bar{u}(x,t) = (W/\sqrt{4\pi t})\,\exp(-x^2/4t)\,.
\end{equation}
However, in stochastic realizations of the KMP model the coarse-grained temperature will fluctuate around the expected profile $\bar{u}(x,t)$, see Fig.~\ref{fig:sim}. To characterize  these non-stationary fluctuations, we will consider the total amount of heat $W_{>}$, observed on the right half line $x>0$ at time $t=T\gg 1$.
The expected value of $W_{>}$ is $W/2$,
and we will study the full time-dependent statistics of the \emph{heat excess}, $J=\int_0^{\infty} u(x,t=T) \,dx -W/2$. Obviously $\mathcal{P}(J,T)$,  the probability distribution of $J$ at time $T$, has a compact support  $|J|\leq W/2$.

Similar non-stationary large-deviation settings, but  with a \emph{step-like} initial condition for the particle density or temperature,  have been recently studied for a whole family of diffusive lattice gases \cite{DG2009b,KrMe,MS2013,MS2014,VMS2014}, of which the KMP model
is an important particular case.  The main working tool of these studies has been the macroscopic fluctuation theory (MFT) \cite{JonaLasinioreview}: a weak-noise theory, whose starting point is fluctuational hydrodynamics (FH) \cite{Spohn,KL,LL}. The FH is a coarse-grained description of the lattice gas, which is accurate
when the characteristic length scale of the problem (here the diffusion length $\sqrt{T}$) and the observation time  $T$ are much larger than the lattice constant $1$ and the inverse elemental rate $1/2$ of the energy exchange, respectively. For diffusive lattice gases with a single conservation law the FH
has the form of a single macroscopic Langevin equation, which accounts for the fluctuational contribution to the heat or mass flux. For the KMP model the Langevin equation reads  \cite{Spohn,KL}
\begin{equation}
    \label{Langevin}
    \partial_t u = \partial_x \left(\partial_x u+\sqrt{2} u \eta\right)\,,
\end{equation}
where $u(x,t)$ is the temperature, and $\eta(x,t)$ is a delta-correlated Gaussian noise: $ \left<\eta(x,t)\right> = 0$ and
$\quad \left<\eta(x,t)\eta(x',t')\right> = \delta(x-x')\delta(t-t')$.

The MFT \cite{JonaLasinioreview}  relies on a saddle-point evaluation of the path integral for the stochastic process, described by Eq.~(\ref{Langevin}). The small parameter of the saddle-point evaluation is again $1/\sqrt{T} \ll 1$: long times correspond to a weak noise. The saddle-point evaluation of the path integral boils down to a minimization of the action functional \cite{SM}, constrained by the specified heat excess $J$ at $t=T$ and obeying the specified initial condition  $u(x,t=0)$.
For the  statistics of the heat (or mass) excess,
the MFT equations and boundary conditions in time were derived in Ref. \cite{DG2009b}, and
we will present them shortly. For completeness, we also present their derivation in \cite{SM}. The solution of the MFT problem describes the \emph{optimal path} of the process: the most likely time history of the temperature  field $u(x,t)$ which dominates the probability distribution $\mathcal{P}(J,T)$ that we are after. The MFT problem, however,  has proven to be very hard to solve analytically, especially for quenched (that is, deterministic) initial conditions \cite{annealed}. In particular, for the KMP model, only small-$J$ \cite{KrMe} and large-$J$ \cite{MS2013} asymptotes 
have been obtained until now (but for a step-like initial condition).

This Letter reports a major advance in this area of statistical mechanics. We present an exact solution to the heat excess statistics problem by uncovering and exploiting complete  integrability of the underlying MFT equations. We obtain explicit results for an initially localized heat pulse, $u (x,t=0) = W \delta(x)$, for which we uncover a nontrivial time-reversal mirror symmetry. These are
the first exact non-steady-state large-deviation results for the statistics of current in a lattice gas of interacting particles for quenched  initial conditions.

\textit{Formulation of the MFT problem} \cite{DG2009b,SM}. -- Let us rescale $t$, $x$ and $u$ by $T$, $\sqrt{T}$ and $W/\sqrt{T}$, respectively. The optimal path we are after is described by two coupled Hamilton's equations for the rescaled temperature field $u(x,t)$ and the conjugate ``momentum density" field $p(x,t)$ which describes the optimal history of the noise $\eta(x,t)$, conditioned on the heat excess $J$.

It is convenient to introduce the (minus) gradient field $v(x,t)=-\partial_x p(x,t)$. In the variables $u$ and $v,$ the MFT equations take the form \cite{DG2009b,MS2013,SM}
\begin{eqnarray}
  \partial_t u &=& \partial_x (\partial_x u+2 u^2 v)\,, \label{d1} \\
  \partial_t v &=& \partial_x (-\partial_{x} v+2 u v^2)\,. \label{d2}
\end{eqnarray}
The rescaled initial condition is
\begin{equation}\label{delta0}
u(x,t=0) = \delta(x)\,.
\end{equation}
The condition on the heat excess  at $t=T$ becomes
\begin{equation}\label{current0}
\int_0^{\infty} u(x,t=1) \,dx -\frac{1}{2} =j  \equiv \frac{J}{W}\,.
\end{equation}
The minimization of the action functional, that enters the constrained path integral, with respect to variations of $u(x,t)$ yields, aside from Eqs.~(\ref{d1}) and~(\ref{d2}),
a second boundary condition in time \cite{DG2009b},
\begin{equation}
\label{vdelta}
v(x,t=1) = -\lambda \,\delta(x)\,,
\end{equation}
where $\lambda$ plays the role of a Lagrange multiplier, to be ultimately fixed by the constraint~(\ref{current0}).

Once $u(x,t)$ and $v(x,t)$ are found, one can calculate
the rescaled action, which can be written as \cite{DG2009b,KrMe,MS2013}
\begin{eqnarray}
  s  = \int_0^1 dt \int_{-\infty}^\infty dx \,u^2 v^2. \label{action0}
\end{eqnarray}
The action yields the probability density ${\cal P}(J,T,W)$ up to a pre-exponent:
\begin{equation}\label{scalingc}
\ln {\mathcal P}(J,T,W)\simeq
-\sqrt{T} \,s\left(\frac{J}{W}\right).
\end{equation}
Since $\sqrt{T} \gg 1$, Eq.~(\ref{scalingc}) has a clear large-deviation structure, and
the action $s$ plays the role of a rate function.

A crucial and previously unappreciated observation is that Eqs. (\ref{d1}) and (\ref{d2})
coincides with the derivative nonlinear Schr\"{o}dinger (DNLS) equation in imaginary time and space \cite{DNLSE}. The DNLS equation (with real time and space) describes propagation of nonlinear electromagnetic waves in plasmas and other media \cite{KN}. An \emph{initial-value} problem for the DNLS equation is completely integrable via the Zakharov-Shabat inverse scattering
method (ISM) adapted by Kaup and Newell for the DNLS \cite{KN}. The MFT formulation presents an difficulty, however, as here one needs to solve a \emph{boundary-value} problem in time, rather than an initial-value problem. Here we overcome this difficulty by (i) making use of a shortcut that allows one to determine the rate function $s(j)$ even without the knowledge of $u(x,t)$ and $v(x,t)$ for all $t$, and
(ii)  exploiting a previously unknown symmetry relation \cite{symmetry}, specific to the initial condition (\ref{delta0}):
\begin{equation}\label{uvsymmetry}
v(x,t) = -\lambda \, u(-x,1-t)\,.
\end{equation}


\textit{Solution of the MFT problem.}--
Equations \eqref{d1} and \eqref{d2} belong to a class of integrable systems for which a Lax pair exists, \textit{i.e.}, as we explain below, the equations are equivalent to the compatibility condition of a system of two linear differential equations. The latter system defines scattering amplitudes which depend on $u$ and $v$. The idea behind the approach that we shall use -- the ISM -- is to consider the time evolution of these scattering amplitudes, which turns out to be very simple, as shown below. By relating these scattering amplitudes, at $t=0$ and $t=1$, to the fields $u$ and $v$, the method will enable us to find the heat excess $j=j(\lambda)$ which suffices for the calculation of $s=s(j)$.

Adapting the derivation of Kaup and Newell \cite{KN} to imaginary time and space, we consider the linear system
\be
\label{psiODEs}
\begin{cases}
\partial_{x}\vect{\psi}(x,t,k)=U(x,t,k)\vect{\psi}(x,t,k)\,,\\
\partial_{t}\vect{\psi}(x,t,k)=V(x,t,k)\vect{\psi}(x,t,k)\,,
\end{cases}
\ee
where $\vect{\psi}(x,t,k)$ is a column vector of dimension 2,
\begin{widetext}
\small
\be
U(x,t,k)=\begin{pmatrix}
- i  k/2 & - i   v\sqrt{  i  k} \\
- i  u\sqrt{ i  k} &   i  k/2 \\
\end{pmatrix}, \;
V(x,t,k)=\begin{pmatrix} k^2/2- i  kuv & - i (\sqrt{ i  k})^3v+ i  \sqrt{ i  k } \,\partial_x v- i  \sqrt{ i  k} 2 v^2u \\
- i (\sqrt{ i  k})^3u+ i  \sqrt{ i  k }\, \partial_x u- i  \sqrt{ i  k} 2 u^2v &   -k^2/2+ i  kuv, \\
\end{pmatrix}  ,
\ee
\end{widetext}
\normalsize
and $k$ is a spectral parameter.
As one can check, the compatibility condition $\partial_{t} \partial_{x}\vect{\psi} = \partial_{x}\partial_{t}\vect{\psi}$, which corresponds to
\begin{equation}
\label{eq:compatibilityUV}
\partial_{t}U-\partial_{x}V+\left[U,V\right] = 0,
\end{equation}
is indeed equivalent to Eqs.~\eqref{d1} and \eqref{d2}.

Let us define the matrix $\mathcal{T}(x,y,t,k)$ as the $x$-propagator of the system \eqref{psiODEs}, namely,  the solution to
\be
\label{eq:dTdx}
\partial_x \mathcal{T}(x,y,t,k)=U(x,t,k)\mathcal{T}(x,y,t,k)
\ee
with $\mathcal{T}(x,x,t,k )=I$ (the identity matrix).
At $x \to \pm \infty$, where the fields $u(x,t)$ and $v(x,t)$ vanish, the matrix $U$ becomes very simple,
\be
U(x\to\pm\infty,t,k)=\begin{pmatrix}-ik/2 & 0\\
0 & ik/2
\end{pmatrix} \,.
\ee
Therefore, it is natural to define the full-space propagator $G(t,k)$ as follows:
\begin{align}
\label{eq:Ttkdef}
G(t,k)&=\lim_{\begin{array}{c}
x\to\infty\\
y\to-\infty
\end{array}}\begin{pmatrix}e^{ i  kx/2} & 0\\
0 & e^{- i  kx/2}
\end{pmatrix}\nn\\&\times \mathcal{T}(x,y,t,k)\begin{pmatrix}e^{- i  ky/2} & 0\\
0 & e^{ i  ky/2}
\end{pmatrix}.
\end{align}
The entries of the matrix $G(t,k)$ are 
the scattering amplitudes of the system \eqref{psiODEs}.
The time evolution of $G(t,k)$ is easy to find. Indeed, the matrix $\mathcal{T}(x,y,t,k)$ satisfies:
\begin{eqnarray}
  \partial_t \mathcal{T}(x,y,t,k) &=& V (x,t,k)\mathcal{T}(x,y,t,k)\nonumber \\
   &-& \mathcal{T}(x,y,t,k)V(y,t,k).
   \label{eq:dTdt}
\end{eqnarray}
One can check that Eq.~\eqref{eq:dTdt} is compatible with \eqref{eq:dTdx} (\textit{i.e.} $\partial_{t} \partial_{x}\mathcal{T} = \partial_{x}\partial_{t}\mathcal{T}$) due to Eq.~\eqref{eq:compatibilityUV}.
The matrix $V(x,t,k)$ too becomes very simple in the limit $x\to\pm\infty$, 
\be
\label{eq:Vlim}
V(x\to\pm\infty,t,k)=\frac{k^{2}}{2}\begin{pmatrix}1 & 0\\
0 & -1
\end{pmatrix} \, .
\ee
Plugging \eqref{eq:Vlim} into \eqref{eq:dTdt}, one finds the time evolution of $\mathcal{T}(x\to \infty, y \to -\infty, t,k)$ which in turn, using \eqref{eq:Ttkdef}, yields that of $G(t,k):$
\begin{align}
\label{eq:TtimeDependence}
G(t,k)=\begin{pmatrix}a(t,k) & \tilde b(t,k) \\
b(t,k) & \tilde a(t,k) \\
\end{pmatrix}=\begin{pmatrix}a(0,k) & \tilde b(0,k)e^{k^2t} \\
b(0,k)e^{-k^2t} & \tilde a(0,k) \\
\end{pmatrix}
\end{align}
where we have introduced here a notation for the matrix elements of $G(t,k)$.

Plugging the temporal boundary conditions \eqref{delta0} and \eqref{vdelta}, we calculate $G(0,k)$ and $G(1,k)$ explicitly by solving Eq.~\eqref{eq:dTdx}, see \cite{SM}. Comparing the two solutions and  using \eqref{eq:TtimeDependence} we obtain
\be
\label{eq:Qpm}
ik\left[Q_{+}(k)+Q_{-}(k)\right]-ikQ_{-}(k)\times ikQ_{+}(k)=-\lambda ike^{-k^{2}}
\ee
where $Q_\pm(k)$ are the Fourier transforms of $v(z,0)$ restricted to $z>0$ and $z<0$, respectively:
\be
\label{Qdef}
Q_-(k)\!=\!\int_{-\infty}^0 \! v(z,0) e^{ i  kz}dz, \;\; Q_+(k)=\int_{0}^\infty \! v(z,0) e^{ i  kz}dz\,.
\ee
We solve Eq.~\eqref{eq:Qpm} in \cite{SM}, with the result
\begin{align}
\label{basically the result}
ikQ_{\pm}(k)&=1-\left(1\pm v_{\pm}\right)e^{\Phi_{\pm}\left(k\right)},\\
\label{Phi}
\Phi_{\pm}\left(k\right)&=\pm\int_{-\infty}^{\infty}\frac{\ln\left(1+i\lambda k'e^{-k'^{2}}\right)}{k'-k\mp i0^{+}}\frac{dk'}{2\pi i} \, ,
\end{align}
where $v_\pm = v(0^\pm, 0)$.

\begin{figure}[ht]
\includegraphics[width=0.3\textwidth,clip=]{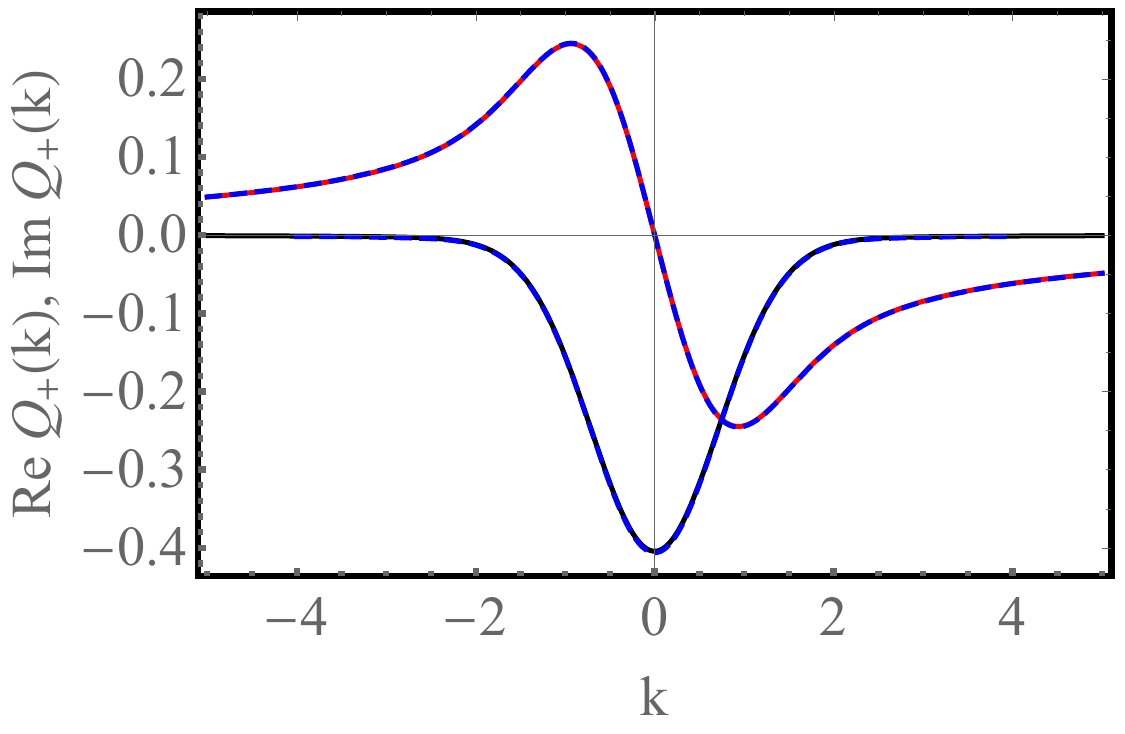}
\caption{Analytical results for $Q_+(k)$, described by Eqs.~(\ref{basically the result}), (\ref{Phi}) and (\ref{vpmfinal}) (solid lines),
versus numerical results (dashed lines) for $\lambda=1$, or $j=0.09568\dots$. The symmetric and antisymmetric curves show $\text{Re}\,Q_+(k)$ and $\text{Im}\,Q_+(k)$, respectively.}
\label{Qplus}
\end{figure}

To compute $v_\pm, $ we  demand that $Q_\pm(k)$ be regular at the origin, corresponding to a vanishing $v(z,0)$ at infinity. Setting $k=0$ in Eq.~\eqref{basically the result} and using the Sokhotski--Plemelj formula 
\be
\int_{-\infty}^{\infty}\frac{f(k)}{k\pm i 0^{+}}\frac{dk}{2\pi i }=\dashint_{-\infty}^{\infty}\frac{f(k)}{k}\frac{dk}{2\pi i }\mp\frac{1}{2}f(0) \,,
\ee
we obtain after some algebra
\begin{align}
\label{vpmfinal}
\pm v_{\pm}=\exp\left[\mp\int_{-\infty}^{\infty}\text{arctan}\left(\lambda k'e^{-k'^{2}}\right)\frac{dk'}{2\pi k'}\right]-1 \, .
\end{align}
Taking the derivative of Eq.~(\ref{basically the result}) with respect to $k$ at $k=0$, yields  \cite{SM}
\begin{equation}
\label{Qplussimpler}
 Q_+(0)=\frac{1}{4\pi} \int_{-\infty}^{\infty} \frac{\ln \left(1+\lambda^2 k^2 e^{-2k^2}\right)}{k^2}\,dk\, - \frac{\lambda}{2} \, .
\end{equation}
Figure \ref{Qplus} shows $\text{Re}\, Q_+(k)$ and $\text{Im}\,Q_+(k)$ versus $k$ at $\lambda=1$, obtained by plugging Eq.~(\ref{vpmfinal}) for $v_{\pm}$ into Eq.~(\ref{basically the result}).
This figure also shows the same quantities computed by solving Eqs.~(\ref{d1}) and (\ref{d2}) numerically with a back-and-forth iteration algorithm \cite{CS}. The analytical and numerical curves are almost indistinguishable.

Using Eqs.~\eqref{current0}, (\ref{uvsymmetry}) and \eqref{Qdef} alongside with the conservation law $\int_{-\infty}^{\infty} u(x,t)\,dx =1$, we determine $j=j(\lambda)$:
\be
\label{jlambda}
j\left(\lambda\right)=\frac{Q_{+}\left(0\right)}{\lambda}+\frac{1}{2}=\frac{1}{4\pi\lambda}\int_{-\infty}^{\infty}\frac{\ln\left(1+\lambda^{2}k^{2}e^{-2k^{2}}\right)}{k^{2}}\,dk \, .
\ee
Now we use a shortcut which makes the results we have obtained so far sufficient for obtaining the rate function $s=s(j)$. The shortcut comes in the form of the relation $ds/dj = \lambda$, which follows from the fact that $j$ and $\lambda$ are conjugate variables, see \textit{e.g.} Ref. \cite{Vivoetal}.
It allows one to calculate $s(j)$ bypassing Eq.~(\ref{action0}) [which would require the knowledge of the whole optimal path $u(x,t)$].
We have
\be
\label{eq:dsdlambda}
\frac{ds}{d\lambda}  =\frac{ds}{dj}\frac{dj}{d\lambda}=\lambda\frac{dj}{d\lambda}
=\frac{dQ_{+}\left(0\right)}{d\lambda}-\frac{Q_{+}\left(0\right)}{\lambda} \, .
\ee
Using Eq.~\eqref{Qplussimpler}, we integrate Eq.~\eqref{eq:dsdlambda} with respect to $\lambda$ to get
\begin{equation}
\label{slambdasimpler}
s(\lambda) = Q_{+}(0)+\int_{-\infty}^{\infty} \frac{\text{Li}_2 \left(-\lambda^2 k^2 e^{-2k^2}\right)}{8\pi k^2}\,dk\, +\frac{\lambda}{2}.
\end{equation}
where
$\text{Li}_{2}\left(z\right)=\sum_{k=1}^{\infty}z^{k}/k^2$ is the dilogarithm function, $Q_{+}(0)$ is given by Eq.~(\ref{Qplussimpler}), and the integration constant was determined from $s(\lambda=0)=0$. Equations~\eqref{jlambda} and \eqref{slambdasimpler} give the complete rate function $s(j)$ in a parametric form and represent the main result of this work.
The exact optimal history of the temperature profile $u(x,t)$ proved difficult to obtain analytically, but it can be computed numerically \cite{SM}.

\begin{figure}[ht]
\includegraphics[width=0.25\textwidth,clip=]{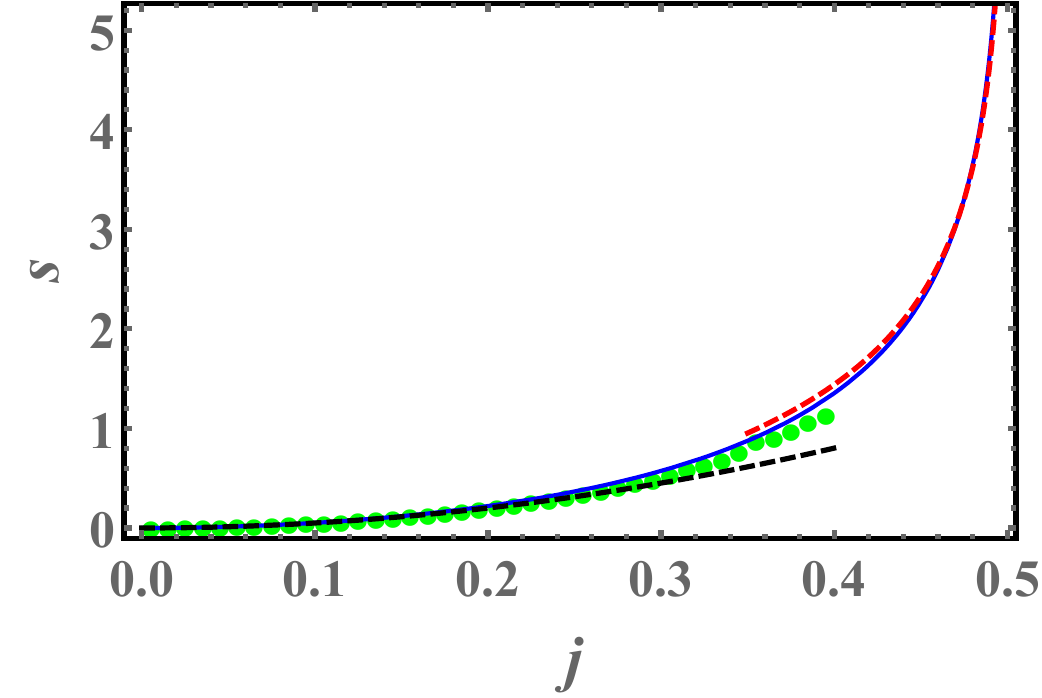}
\caption{The exact rate function $s(j)$, given  by Eqs.~(\ref{jlambda}) and \eqref{slambdasimpler}
 (solid line) and two asymptotes: $s(|j|\ll 1)=\sqrt{8\pi} j^2$ and~Eq.~(\ref{slargej}) (dashed lines). Symbols: properly rescaled data from $10^6$ direct Monte-Carlo simulations of the microscopic KMP model for 
 $T=10^2$, see \cite{SM} for details.}
\label{ratefunction}
\end{figure}

Figure \ref{ratefunction} shows $s(j)$ alongside with two asymptotes: $j \to 0$ and $|j|\to 1/2$, which correspond to $\lambda \to 0$ and $|\lambda|\to \infty$, respectively. Also shown are results of Monte-Carlo simulations. The asymptote $\lambda\to 0$ can be obtained either from the exact rate function \eqref{jlambda} and \eqref{slambdasimpler} \cite{SM}, or from a perturbative expansion applied  directly to the MFT equations \cite{KrMe}. By virtue of the symmetry (\ref{uvsymmetry}), the latter can be done very easily. Indeed, in the leading order in $\lambda\ll 1$ Eqs.~(\ref{action0}), (\ref{uvsymmetry}) and  (\ref{meanfield}) yield
\begin{equation}\label{actionsimple}
s(\lambda) \!\simeq\! \lambda^2 \!\int_0^1 \! dt \int_{-\infty}^{\infty} \! dx\, \bar{u}^2(x,t)\,\bar{u}^2(-x,1-t)
=\frac{\lambda ^2}{8 \sqrt{2 \pi}}\,.
\end{equation}
The shortcut relation $ds/dj = \lambda$ can be rewritten as $(ds/d\lambda) (d\lambda/dj) = \lambda$. Combined with Eq.~(\ref{actionsimple}) it yields $s(j\to 0) \simeq \sqrt{8\pi} j^2$. Then, from Eq.~(\ref{scalingc}), we see that \emph{typical} fluctuations of $J$ are normally distributed with variance $W^2/(32 \pi T)^{1/2}$. The $T^{-1/2}$ scaling of the variance should be contrasted with the $T^{1/2}$ scaling, obtained for a step-like  initial condition \cite{DG2009b,DG2009a,KrMe}. However, the \emph{relative} magnitude of the fluctuations -- the ratio of the standard deviation and the average transferred heat -- has the same scaling $T^{-1/4}\ll 1$ in both settings, as to be expected from the law of large numbers.

The asymptote of $|\lambda| \to \infty$ is more subtle \cite{SM}. The final result, already in terms of $j$, is
\begin{align}
\label{slargej}
&s\left(|j|\to 1/2\right)\simeq\frac{4\left[-\frac{1}{2}W_{-1}\left(-\frac{1}{2}\pi^{2}\Delta^{2}\right)\right]^{3/2}}{3\pi}\nn\\
&=\frac{4}{3\pi}\ln^{3/2}\left(\frac{2}{\pi\Delta}\,\sqrt{\ln\frac{2}{\pi\Delta}\sqrt{\ln\frac{2}{\pi\Delta}\dots}}\right)\,,
\end{align}
where $\Delta \equiv 1/2-|j| \ll 1$, and $W_{-1}(\dots)$ is the proper branch of the product log (Lambert $W$) function \cite{LambertW}. At $j=1/2$, $s$ diverges and  $\mathcal{P}$ vanishes, as to be expected. Nested-log large-current asymptotes similar to Eq. (\ref{slargej}) appear to be typical for the KMP model and other models of the hyperbolic universality class \cite{MS2013,MS2014,ZarfatyM}.

\textit{Discussion.}--
By combining the MFT and the ISM, we calculated exactly the rate function $s(j)$, see Eqs.~\eqref{jlambda} and \eqref{slambdasimpler}, which describes the full long-time statistics of nonstationary heat transfer in the KMP model for an initially localized heat pulse. This is
the first exact non-steady-state large-deviation result for the statistics of current in a lattice gas of interacting particles for quenched  initial conditions.
It opens the way to extensions of the ISM to additional fluctuating quantities of the KMP model. Another challenging goal is to apply the ISM to the simple symmetric exclusion process (SSEP)  \cite{Spohn,Liggett,KL,Krapivskybook} -- a lattice-gas model with quite different properties \cite{MS2014}. Encouragingly, the MFT equations for the SSEP (see \textit{e.g.} \cite{DG2009b}) can be mapped to Eqs.~(\ref{d1}) and (\ref{d2}) via a canonical transformation. This transformation, however, complicates the boundary conditions in time.

From a more general perspective, the MFT of lattice gases is a particular case of the weak-noise theory, or optimal fluctuation method (OFM): a highly versatile framework which captures a broad class of large deviations in macroscopic systems.  For non-stationary processes the OFM equations -- coupled nonlinear partial differential equations for the optimal path -- are usually very hard to solve exactly. One class of problems of this type, which has received much recent attention, deals with the
complete one-point height statistics of an interface whose dynamics is described by the Kardar-Parisi-Zhang (KPZ) equation \cite{KPZ}. The OFM captures the complete KPZ height statistics at short times \cite{KK,MKV,KMS,JKM,Lin,Lamarre}. Here too, a previous analytical progress in the solution of the OFM equations was limited to asymptotics of very large or very small interface height. But very recently these OFM equations  -- which coincide with the Nonlinear Schr\"{o}dinger equation (NLS) (not the derivative one) \cite{JKM} -- have been solved
exactly \cite{KLD1,KLD2} by the ISM for several ``standard" initial conditions. The two integrable systems, the NLS and DNLS, are closely related, so our approach can be compared with that of Refs. \cite{KLD1,KLD2}. We used only standard techniques of the ISM which do not rely on additional tools, such as Fredholm determinants used in Refs. \cite{KLD1,KLD2}. Because of its relative simplicity our approach appears to be more readily adaptable to solving additional large-deviation problems \cite{BSM2}.

\textit{Acknowledgments.} -- The research of E.B. and B.M. is supported by the Israel Science Foundation (grants No. 1466/15 and 1499/20, respectively). N.R.S. acknowledges
support from the Yad Hanadiv fund (Rothschild
fellowship).

\begin{widetext}

\newpage

\setcounter{secnumdepth}{2}

\newcommand{\eqrefMT}[1]{(#1)}
\newcommand{\Langevin}{2}
\newcommand{\dOne}{3} 
\newcommand{\dTwo}{4} 
\newcommand{\vdelta}{7}
\newcommand{\actionZero}{8}
\newcommand{\eqdTdx}{14}
\newcommand{\eqTtkdef}{16}
\newcommand{\eqTtimeDependence}{19}
\newcommand{\eqQpm}{20} 
\newcommand{\Qdef}{21}
\newcommand{\basicallytheresult}{22}
\newcommand{\eqPhi}{23}
\newcommand{\Qplussimpler}{26}
\newcommand{\jlambda}{27}
\newcommand{\eqdsdlambda}{28}
\newcommand{\slambdasimpler}{29}
\newcommand{\slargej}{31}

\begin{large}
\begin{center}
Supplementary Material for\\  ``Inverse Scattering Method Solves the Problem of Full Statistics of Nonstationary Heat Transfer in the Kipnis-Marchioro-Presutti Model" by E. Bettelheim \textit{et al}.
\end{center}
\end{large}

\bigskip
Here we give some technical details of the calculations described in the main text of the Letter.
\bigskip


\renewcommand{\theequation}{S\arabic{equation}}
\setcounter{equation}{0}

\section{Derivation of the MFT equations and boundary conditions}

We begin by defining an auxiliary potential
\be
\psi\left(x,t\right)=\int_{-\infty}^{x}u\left(y,t\right)dy \, .
\ee
Integrating Eq.~\eqref{Langevin} of the main text with respect to $x$ and using the boundary condition $u(x\to-\infty,t)\to0$, we obtain a Langevin equation for $\psi(x,t)$:
\be
\label{psiLangevin}
\partial_{t}\psi=\partial_{x}^{2}\psi+\sqrt{2}\,\partial_{x}\psi\,\eta\,.
\ee
Now we use a path-integral approach, by writing the probability density functional $P[\eta]$ of the white Gaussian noise term $\eta(x,t)$:
\be
P\left[\eta\right]\sim\exp\left(-\int_{0}^{T}dt\int_{-\infty}^{\infty}dx\,\frac{\eta^{2}}{2}\right)\,.
\ee
We now express $\eta$ through $\psi$ via Eq.~\eqref{psiLangevin}, enabling us to write the probability density for a given history of the system:
\be
\label{Spsi}
-\ln P\left[\psi\right]\simeq S\left[\psi\right]\equiv\frac{1}{4}\int_{0}^{T}dt\int_{-\infty}^{\infty}dx\,\left(\frac{\partial_{t}\psi
-\partial_{x}^{2}\psi}{\partial_{x}\psi}\right)^{2}\,,
\ee
where $S\left[\psi\right]$ is the action functional.
The presence of the large parameter $\sqrt{T}\gg 1$ enables us to calculate $\mathcal{P}(J,T)$ via a saddle-point evaluation of the path integral. Within this framework, $-\ln\mathcal{P}\left(J,T\right)$ is given by minimum of the action functional $S$, constrained on the initial condition $u(x,0)=W\delta(x)$ and on a given value of heat excess $J$.
The heat excess, which we recall is $J=\int_0^{\infty} u(x,t=T) \,dx -W/2$, is conveniently rewritten as
$J=W/2-\psi(0,T)$ (where we used the conservation law $\int_{-\infty}^{\infty} u(x,t=T) \,dx = W$).
To take into account the latter constraint on $J$, we add the term $-\Lambda\psi\left(0,T\right)$ to the action, where $\Lambda$ is a Lagrange multiplier, i.e., we minimize the constrained functional
\be
S_{\Lambda}\left[\psi\right]=S\left[\psi\right]-\Lambda\psi\left(0,T\right)=\frac{1}{4}\int_{0}^{T}dt\int_{-\infty}^{\infty}dx\,\left(\frac{\partial_{t}\psi-\partial_{x}^{2}\psi}{\partial_{x}\psi}\right)^{2}-\Lambda\psi\left(0,T\right) \, .
\ee

The subsequent procedure is pretty standard. Consider a variation $\psi(x,t) \to \psi(x,t)+\delta \psi(x,t)$. This leads to a variation of $S_{\Lambda}\left[\psi\right]$, which, to first order in $\delta \psi$ is given by
\bea
\label{deltaS1}
\delta S_{\Lambda} &=&S_{\Lambda}\left[\psi+\delta\psi\right]-S_{\Lambda}\left[\psi\right]=\frac{1}{2}\int_{0}^{T}dt\int_{-\infty}^{\infty}dx\,\left(\frac{\partial_{x}^{2}\psi-\partial_{t}\psi}{\partial_{x}\psi}\right)\left[\frac{\partial_{x}^{2}\delta\psi-\partial_{t}\delta\psi}{\partial_{x}\psi}-\frac{\partial_{x}^{2}\psi-\partial_{t}\psi}{\left(\partial_{x}\psi\right)^{2}}\partial_{x}\delta\psi\right]-\Lambda\delta\psi\left(0,T\right)\nn\\
&=& \int_{0}^{T}dt\int_{-\infty}^{\infty}dx\,\partial_{x}p\left(\partial_{x}^{2}\delta\psi-\partial_{t}\delta\psi-2\partial_{x}p\partial_{x}\psi\partial_{x}\delta\psi\right)-\Lambda\delta\psi\left(0,T\right)\,,
\eea
where we have defined the momentum density gradient
\be
\label{pdef}
\partial_{x}p=\frac{\partial_{x}^{2}\psi-\partial_{t}\psi}{2\left(\partial_{x}\psi\right)^{2}} \; .
\ee
Integrating by parts in Eq.~\eqref{deltaS1}, we obtain
\bea
\label{deltaS2}
\delta S_{\Lambda}&=&\int_{0}^{T}dt\int_{-\infty}^{\infty}dx\,\left\{ \partial_{x}^{3}p+\partial_{xt}p+2\partial_{x}\left[\left(\partial_{x}p\right)^{2}\partial_{x}\psi\right]\right\} \delta\psi \nn\\
&+&\int_{-\infty}^{\infty}dx\,\left[\partial_{x}p\left(x,0\right)\delta\psi\left(x,0\right)-\partial_{x}p\left(x,T\right)\delta\psi\left(x,T\right)-\Lambda\delta\left(x\right)\delta\psi\left(x,T\right)\right]
\eea
where the first two terms in the single integral are the boundary terms originating from the integration by parts in time.
For the quenched (deterministic) initial condition, the variation $\delta\psi\left(x,0\right)$ vanishes.

The second MFT equation, Eq.~\eqref{d2} in the main text is now obtained by requiring the double integral in \eqref{deltaS2} to vanish for arbitrary $\delta \psi$ (recalling that $v=-\partial_{x}p$ and $u=\partial_{x} \psi$). The first MFT equation, Eq.~\eqref{d1} in the main text, follows from Eq.~\eqref{pdef} after multiplying by the denominator and then taking a spatial derivative.
Note that, under the rescalings of $x$, $t$ and $u$ described in the text, $v$ should be rescaled by $1/W$. These rescalings leave the MFT equations invariant.
%
Requiring the single integral in Eq.~\eqref{deltaS2} to vanish for arbitrary $\delta\psi\left(x,T\right)$, we obtain the boundary condition
\be
\partial_{x}p\left(x,T\right) = -\Lambda\delta(x) \, .
\ee
After the rescaling, this becomes Eq.~\eqref{vdelta} in the main text, where $\lambda = W\Lambda$ is a rescaled Lagrange multiplier.
Finally, using Eq.~\eqref{pdef} in \eqref{Spsi} we find that the action can be rewritten as $S=\int_{0}^{T}dt\int_{-\infty}^{\infty}dx\,u^{2}v^{2}$ which, after the rescaling, leads to $S=\sqrt{T} \,  s$ where the rescaled action $s$ is given by Eq.~\eqref{action0} in the main text.
The large parameter $S\sim\sqrt{T}\gg1$ justifies \textit{a posteriori} the saddle-point approximation that we used.

\section{Solving the scattering problem at $t=0$ and $t=1$}

Let us find the matrix $\mathcal{T}(x,y,0,k) $ at $t=0$. By solving Eq.~\eqref{eq:dTdx} of the main text at $t=0$, using  $u(x,0)=\delta(x)$, one gets
\be
\label{eq:Txyt0}
\mathcal{T}(x,y,0,k)=  \begin{cases}
\begin{pmatrix}e^{- i  k(x-y)/2} & - i \sqrt{ i  k/2}e^{- i  k(x-y)/2}I_{v}(x,y)\\
0 & e^{ i  k(x-y)/2}
\end{pmatrix}\,, &  xy>0\,,\\[5mm]
\begin{pmatrix}e^{ i  k(y-x)/2}\left[1\pm i  kI_{v}(x,0)\right] & - i \sqrt{ i  k}e^{ i  k(y-x)/2}\left[I_{u}(x,y)\pm i  kI_{u}(0,y)I_{u}(x,0)\right]\\
\pm i \sqrt{ i  k}e^{ i  k(x+y)/2} & e^{ i  k(x-y)/2}\pm i  ke^{ i  k(x+y)/2}I_{u}(0,y)
\end{pmatrix}\,, &  xy<0\,,
\end{cases}
\ee
 where
 \begin{align}
 I_v(x,y)=\int_y^x v(z) e^{ i  k(z-y)}dz, \qquad I_{u}(x,y)=\int_{y}^{x}u(z,1)e^{- i  k(z-y)}dz,
 \end{align}
and in the second case in \eqref{eq:Txyt0}, the sign $\pm $ is to be taken as the sign of $y.$
Plugging \eqref{eq:Txyt0} into Eq.~\eqref{eq:Ttkdef} of the main text, we compute $G(0,k)$:
\be
\label{T(t=0)}
G(0,k)=\begin{pmatrix}1- i  kQ_{+}(k) & - i \sqrt{ i  k}\left[Q(k)- i  kQ_{-}(k)Q_{+}(k)\right]\\
- i \sqrt{ i  k} & 1- i  kQ_{-}(k)
\end{pmatrix}
\ee
in terms of $Q_{\pm}(k)$ which are defined in Eq.~\eqref{Qdef} in the main text, with $Q(k) = Q_+(k) + Q_-(k)$.

It is useful to compare this result to the one obtained at $t=1$.
Here we have $v(x,1)=  - \lambda\delta(x)$.  Similarly to the $t=0$ case, one gets
\be
\mathcal{T}(x,y,1,k)= \begin{cases}
\begin{pmatrix}e^{- i  k(x-y)/2} & 0\\
- i  \sqrt{ i  k} \,  e^{ i  k(x-y)k/2}I_{u}(x,y) & e^{ i  k(x-y)/2}
\end{pmatrix}\,, & xy>0\,,\\[5mm]
\begin{pmatrix}e^{- i  k(x-y)/2}\pm\lambda i  ke^{ i  k(x+y)/2}I_{u}(0,y) & \pm i \lambda\sqrt{ i  k}e^{- i  k(x+y)/2}\\
- i \sqrt{ i  k}e^{ i  k(x-y)/2}\left[I_{u}(x,y)\pm\lambda i  kI_{u}(0,y)I_{u}(x,0)\right] & e^{ i  k(x-y)/2}\left[1\pm\lambda i  kI_{u}(x,0)\right]
\end{pmatrix}\,, & xy<0\,,
\end{cases}
\ee
 where in the second case, the sign $\pm $ is to be taken as the opposite of the sign of $y.$
 Now compute $G(1,k)$:
\be
 \label{T(t=1)}
G(1,k)=\begin{pmatrix}1+\lambda i  kR_{+}(k) & + i \lambda\sqrt{ i  k}\\
- i \sqrt{ i  k}\left[R(k)+\lambda i  kR_{-}(k)R_{+}(k)\right] & 1+\lambda i  kR_{-}(k)
\end{pmatrix}
\ee
where
\be
R_+(k)=\int_{-\infty}^0 u(z,1) e^{- i  kz}dz, \quad R_-(k)=\int_{0}^\infty u(z,1) e^{- i  kz}dz, \quad R(k)=R_+(k)+R_-(k) \, .
\ee
Comparing the upper-right elements of $G(0,k)$ from Eq.~(\ref{T(t=0)}) and of $G(1,k)$ from Eq.~(\ref{T(t=1)}), using Eq.~\eqref{eq:TtimeDependence} of the main text, leads to Eq.~\eqref{eq:Qpm} of the main text.

\section{Solving Eq.~(\ref{eq:Qpm}) of the main text}

We can complete the squares in Eq.~\eqref{eq:Qpm} of the main text by writing
\begin{align}
\label{eq:QgivenM}
 i  kQ_\pm(k)=1-(1\pm v_{\pm})e^{M_\pm(k)  }\,,
\end{align}
where $v_\pm = v(0^\pm, 0)$.
This turns Eq.~\eqref{eq:Qpm} of the main text into
\begin{align}
\label{eq:vpmMpm}
\left(1+v_+\right)\left(1-v_-\right)e^{M_+(k)+M_-(k)}=1 + i \lambda    k   e^{-k^2},
\end{align}
which has the solution
\be
\label{eq:Mpmsol}
M_{\pm}(k)= \pm\int_{-\infty}^{\infty}\frac{\ln\left(1 +  i \lambda k'e^{-k'^{2}}\right)}{k'-k\mp i 0^{+}}\frac{dk'}{2\pi i }
\ee
provided the condition
\begin{align}
\left(1+v_+\right)\left(1-v_-\right)=1\label{q+q-firstEq.}
\end{align}
is satisfied.
Eq. \eqref{eq:Mpmsol} is derived by  noting that $Q_\pm$ are analytic in the upper and lower half plane respectively and are well-behaved when $k$ is allowed to reach infinity through the respective half-planes. We then use the well-known decomposition $f(k) = f_+(k) + f_-(k)$ of a general function $f(k)$ into functions analytic in the upper and lower half-planes, $f_\pm(k)$, respectively, given by $f_\pm(k)=\int \frac{f(k')}{k'-k\mp i 0^+}\frac{dk'}{2\pi i}$. This decomposition is applied to the logarithm of Eq. (\ref{eq:vpmMpm}). Plugging Eq.~\eqref{eq:Mpmsol} into Eq.~\eqref{eq:QgivenM}, we obtain the solution given in Eqs.~\eqref{basically the result} and \eqref{Phi} of the main text.

\section{Calculating $Q_+(0)$}

Taking the derivative of Eq.~\eqref{basically the result} with respect to $k$ at $k=0$, yields
 the equation:
 \begin{align}
 Q_{+}(0)&= i  \left(1+ v_{+}\right)\frac{d}{dk} \exp\left[\dashint_{-\infty}^{\infty}\frac{\ln\left(1+ i \lambda k'e^{-k'^{2}}\right)}{k'-k}\frac{dk'}{2\pi i }+\frac{1}{2}\ln\left(1+ i \lambda ke^{-k^{2}}\right)\right]_{k=0}\nn\\
&= i \frac{d}{dk}\left[\dashint_{-\infty}^{\infty}\frac{\ln\left(1+ i \lambda k'e^{-k'^{2}}\right)}{k'-k}\frac{dk'}{2\pi i }+\frac{1}{2}\ln\left(1+ i \lambda ke^{-k^{2}}\right)\right]_{k=0}=\nn\\
&=\dashint_{-\infty}^{\infty}\frac{\ln\left(1 +  i \lambda k'e^{-k'^{2}}\right)}{k'^{2}}\frac{dk'}{2\pi} -\frac{\lambda}{2}
\end{align}
The imaginary part of the integrand is an odd function of $k$ and therefore does not contribute to the integral. Keeping only the real part and simplifying it,
we obtain \eqref{Qplussimpler} of the main text,
where we replaced the principal value integral by a regular integral since the integrand is regular at $k=0$.

\section{Asymptotics of small and large $\lambda$}

At small $\lambda$ Eq.~\eqref{Qplussimpler} of the main text yields
\begin{equation}\label{jLT}
Q_{+}(0)(\lambda\ll1)\simeq\frac{1}{4\pi}\int_{-\infty}^{\infty}\frac{\lambda^{2}k^{2}e^{-2k^{2}}}{k^{2}}\,dk-\frac{\lambda}{2}=-\frac{\lambda}{2}+\frac{\lambda^{2}}{4\sqrt{2\pi}}\,,
\end{equation}
therefore $j(\lambda\ll 1) \simeq \lambda/(4\sqrt{2\pi})$. Now, using the $|z|\ll 1$ asymptotic
$\text{Li}_2(-z) \simeq -z$ in the integrand of Eq.~\eqref{slambdasimpler} of the main text, we obtain  after a simple algebra: $s(\lambda\ll 1) \simeq \lambda^2/(8\sqrt{2\pi})$. This yields the asymptotic behavior
$s(j \ll 1) \simeq 2\sqrt{2\pi} \, j^2$ given in the main text.

Now we consider the $|\lambda| \gg 1$ asymptote and start from Eq.~\eqref{Qplussimpler} of the main text. Due to the symmetry $j(-\lambda) = - j(\lambda)$, $s(-\lambda) = s(\lambda)$ we consider only $\lambda>0$. Let us denote the integrand in Eq.~\eqref{Qplussimpler} (including the factor $1/4\pi$) by
$$
F(\lambda, k) =\frac{\ln \left(1+\lambda^2 k^2 e^{-2 k^2}\right)}{4\pi k^2}\,.
$$
We can recast it as
\begin{equation}\label{b2}
F(\lambda, k) =\frac{\ln \left(1+\lambda^2 k^2\right)}{4\pi k^2}+\Phi(\lambda,k)\,,
\end{equation}
where
\begin{equation}\label{b3}
\Phi(\lambda, k) =\frac{\ln \left(\frac{1+\lambda^2 k^2 e^{-2 k^2}}{1+\lambda^2 k^2}\right)}{4\pi k^2}\,.
\end{equation}
The integral over the first term of Eq.~(\ref{b2}) yields $\lambda/2$, and we now focus on the integral of $\Phi(\lambda,k)$. For $\lambda \gg 1$, $\Phi(\lambda,k)$ as a function of $k$ behaves as follows (see Fig. \ref{FigPhi}). At $1\ll \lambda k < \sqrt{\ln \lambda}$, $\Phi(\lambda,k)$ is approximately constant and equal to $-1/(2\pi)$. This asymptote is obtained when
neglecting $1$ in the numerator and in the denominator of the fraction inside the logarithm in Eq.~(\ref{b3}). For $k\gtrsim \sqrt{\ln \lambda}$, $\Phi(\lambda,k)$ behaves as
\begin{equation}\label{b4}
\Phi(\lambda,k)\simeq \frac{\ln \left(\frac{1}{\lambda^2 k^2}\right)}{4 \pi  k^2}\,.
\end{equation}
This asymptote is obtained when neglecting the second term in the numerator and $1$ in the denominator of the fraction inside the logarithm in Eq.~(\ref{b3}). Importantly, the transition from the asymptote  $-1/(2\pi)$ to the asymptote (\ref{b4}) occurs in a narrow boundary layer around $k=\sqrt{\ln \lambda}$, whose width goes to zero as $\lambda \to \infty$. Furthermore, the region of $k\lesssim 1/\lambda$ contributes a term $O(1/\lambda)$ to the integral which, as we shall see \textit{a posteriori}, is negligible.  Therefore, we can divide the integration region into two subregions: $0<k<\sqrt{\ln \lambda}$ and $\sqrt{\ln \lambda}<k<\infty$, and use the asymptote $\Phi(\lambda,k)\simeq -1/(2\pi)$ in the former subregion, and Eq.~(\ref{b4}) in the latter one. Keeping the leading and two subleading terms in the result and multiplying it by 2 to account for $k<0$, we obtain
\begin{equation}\label{Qlarge}
Q_+(0) = -\frac{2 \sqrt{\ln \lambda
   }}{\pi }-\frac{\ln \ln \lambda }{2 \pi  \sqrt{\ln
   \lambda }}-\frac{1}{\pi  \sqrt{\ln \lambda}}+ \dots
\end{equation}

\begin{figure*}[ht]
\centering
\includegraphics[angle=0,width=0.4\textwidth]{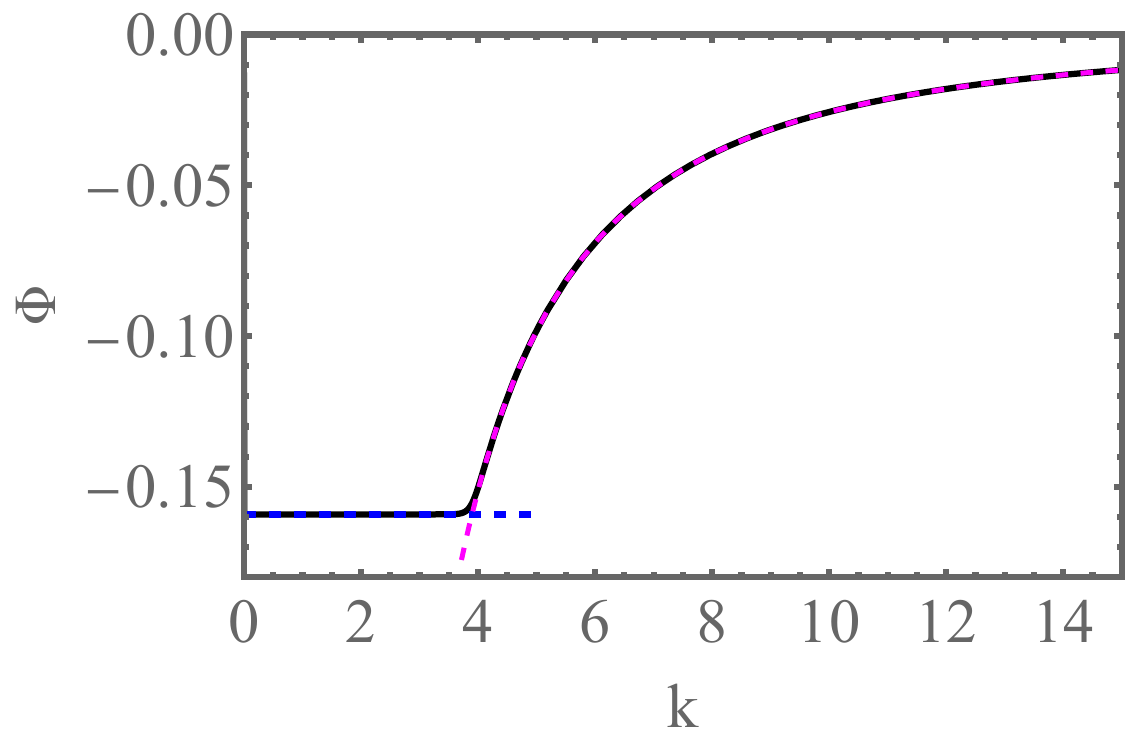}
\caption{Exact $\Phi(\lambda,k)$, given by Eq.~(\ref{b3}) (black solid line), and the asymptotes $\Phi(\lambda,k) \simeq -1/(2\pi)$ and Eq.~(\ref{b4}) (blue and magenta dashed lines, respectively), for $\lambda=10^6$. The region of $k\lesssim 1/\lambda$ cannot be seen on this scale, and its contribution to the integral is negligible.}
\label{FigPhi}
\end{figure*}

Using the relation \eqref{jlambda} of the main text between $j$ and $Q_+(0)$, we obtain in the leading order:
\begin{equation}\label{jlargelambda}
j(\lambda\gg 1) \simeq \frac{1}{2}-\frac{2 \sqrt{\ln \lambda }}{\pi \lambda}.
\end{equation}
To calculate the large-$\lambda$ asymptote of $s(\lambda)$, we plug \eqref{jlargelambda} into the first equality in \eqref{eq:dsdlambda} in the main text to get
\be
\frac{ds}{d\lambda}=\lambda\frac{dj}{d\lambda}\simeq\frac{2\sqrt{\ln\lambda}}{\pi\lambda}
\ee
Integrating this over $\lambda$ we obtain
\begin{equation}\label{slargelambda}
s(\lambda\gg 1) \simeq \frac{4}{3\pi} (\ln \lambda)^{3/2}.
\end{equation}
(Note that, in the limit $\lambda \gg 1$ that we consider here, the integration constant is not important.) Solving Eq. (\ref{jlargelambda}) for $\lambda$, we obtain
\begin{equation}\label{lambdalargejexact}
\lambda\left(j\to\left(\frac{1}{2}\right)^{-}\right)\simeq e^{-\frac{1}{2}W_{-1}\left(-\frac{1}{2}\pi^{2}\Delta^{2}\right)},
\end{equation}
where $\Delta \equiv 1/2-j \ll 1$, and $W_{-1}(\dots)$ is the proper branch of the product log (Lambert $W$) function \cite{LambertWSM}.
Plugging  Eq.~(\ref{lambdalargejexact}) into Eq.~(\ref{slargelambda}), we obtain the large-$j$ asymptote \eqref{slargej} of the main text.

\section{Obtaining the optimal path $u(x,t)$ numerically at all times}

Our exact solution gives the rate function $s(j)$ but it does not give the optimal path $u(x,t)$ at all times. To calculate the latter analytically, one would need to solve Eq.~\eqref{eq:dTdx} of the main text at arbitrary times $0\leq t\leq 1$, and this is far more challenging than solving it only at $t=0$ and $t=1$, as we did.
The optimal path can, however, be computed numerically, using the back-and-forth iteration algorithm due to Chernykh and Stepanov \cite{CSSM}. The results of one such calculation are shown in Fig.~\ref{fig:uxt}.
\begin{figure}[ht]
\includegraphics[width=0.3\textwidth,clip=]{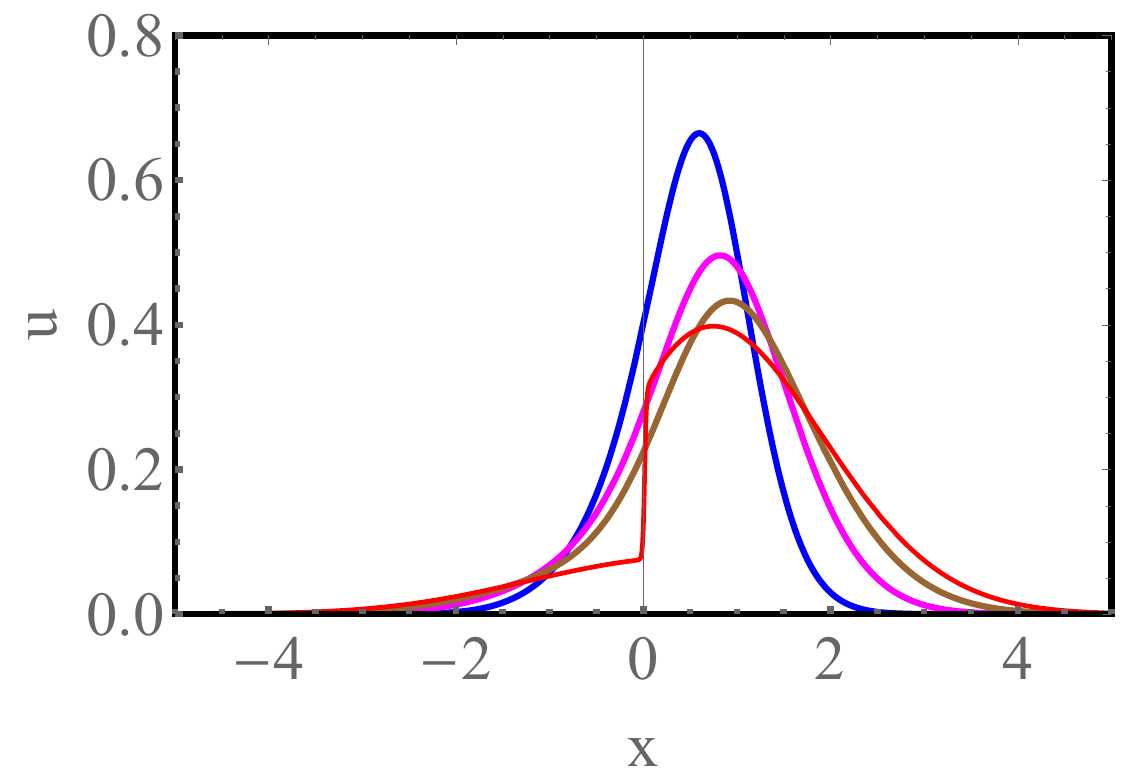}
\caption{The optimal temperature profile $u(x,t)$ for $\lambda=10$ (corresponding to $j \simeq 0.38$) at times $1/4$, $1/2$, $3/4$ and $1$. Noticeable is a shock-like singularity of $u$ at $x=0$ and $t=1$.}
\label{fig:uxt}
\end{figure}

\section{Monte-Carlo simulations}

Here we describe how the numerical data in Fig.~\ref{ratefunction} of the main text was generated.
We performed Monte-Carlo (MC) simulations of the microscopic KMP model.
The dynamical object in the simulations is the vector of energies at $2L+1$ lattice sites,
$u_{-L},\dots,u_{-1},u_{0},u_{1},\dots u_{L}$,
where the site $0$ represents the origin, $x=0$.
Initially, all of the energy is at the origin, $u_{i}=\delta_{i,0}$ (this corresponds to $W=1$).
At every Monte-Carlo step, we randomly choose one of the $2L$ adjacent pairs of lattice sites $\left(i,i+1\right)$, and randomly redistribute their total energy between them, i.e.,
$u_{i},u_{i+1}\to\tilde{u}_{i},\tilde{u}_{i+1}$ where $\tilde{u}_{i}$ is sampled from a uniform distribution between $0$ and $u_{i}+u_{i+1}$, and $\tilde{u}_{i+1}=u_{i}+u_{i+1}-\tilde{u}_{i}$.
Since (in our convention) the microscopic rate of this elementary process is $2$, the total number of steps in the simulation is randomly sampled from a Poisson distribution with mean $4TL$.
The excess heat was measured at $t=T$ as
$J=u_{0}/2+\sum_{i=1}^{L}u_{i} - 1/2$.
Since the distribution is exactly symmetric, $\mathcal{P}(J,T) = \mathcal{P}(-J,T)$, a histogram of the absolute values of the simulated $J$'s was constructed.
For the data plotted on Fig.~\ref{ratefunction} of the main text, the parameters used were $T=100$ and $L=25$ (we found that increasing $L$ beyond this value did not noticeably affect the results displayed in the figure). The symbols in the figure correspond to
\be
\label{MCNormalization}
-\ln\left(\sqrt{2\pi V}\,\mathcal{P}\left(J,T\right)\right)/\sqrt{T}\,,
\ee
where the probability density function $\mathcal{P}(J,T)$ was computed from the histogram of the simulated $J$'s, and $V=1/\sqrt{32\pi T}$ is our prediction for the variance [see the paragraph following Eq.~\eqref{actionsimple} of the main text].
The term $\sqrt{2\pi V}$ in Eq.~\eqref{MCNormalization} comes from the normalization of the central part of the distribution, rather than from a systematic calculation of the pre-exponential factor in
$\mathcal{P}(J,T)$ which is beyond the leading-order MFT.  
The normalization approximation of the pre-exponential factor introduces a small systematic error at moderate $J$ which is discernible in Fig.~\ref{ratefunction} of the main text. The relative error, introduced by the pre-exponential factor, must go to zero as $\sqrt{T}$ goes to infinity. 

As to be expected, direct MC simulations become prohibitively long for larger $J$ and $T$. Special methods of large deviation sampling should be used for these purposes.

\end{widetext}

\end{document}